# TESTS OF THE CERN PROTON LINAC PERFORMANCE FOR LHC-TYPE BEAMS

C.E. Hill, A. Lombardi, R. Scrivens, M. Vretenar
CERN, CH-1211 Geneva 23, Switzerland
A. Feschenko, A. Liou
Institute for Nuclear Research, Moscow, 117312 Russia

*Abstract*

As the pre-injector of the LHC injector chain, the proton linac at CERN is required to provide a high-intensity (180mA) beam to the Proton Synchrotron Booster. The results of measurements at this intensity will be presented. Furthermore, the linac is now equipped with bunch shape monitors from INR, Moscow, which have allowed the comparison of the Alvarez tank RF settings with simulations.

## 1 INTRODUCTION

Linac 2 has been in operation since September 1978 and routinely supplies protons during 6700 hours of operation per year. The machine consists of a Duoplasmatron proton source at 90 kV, a 750 keV RFQ, and three Alvarez tanks accelerating the beam to 50 MeV. In normal operation a 170 mA proton beam is injected into the Proton Synchrotron Booster (PSB) with a pulse length up to 150 µs, at a repetition rate of 0.8 Hz [1].

From 2005 onwards, the linac will function as the pre-injector for the Large Hadron Collider (LHC), for which 180 mA is desired [2]. This is difficult to achieve because the longitudinal beam dynamics are strongly space-charge limited at the low energy end of tank 1.

Within the framework of upgrading towards this intensity the 750 kV Cockcroft-Walton was replaced with an RFQ in 1993. More recently, three Bunch Shape Monitors (BSMs) have been installed, to allow the study of the beam dynamics of the Alvarez tanks.

## 2 HIGH CURRENT TESTS

The optimisation of the linac to produce higher output currents was performed during 1999. A comparison of the readings of the current transformers along the length of the linac, for the high current case and during normal operation, is given in Figure 1. It is clear that the principle gain in this case was the 32% higher current from the source. From this higher current, the improvement of 10% was possible at the end of the linac, and the greater losses in the transfer line (after the linac) led to a final improvement at the entrance of the PSB injection line of 6%.

The large losses from TRA02 to TRA06 are mostly due to the loss of the $H_2^+$ beam at the entrance of the RFQ.

The beam parameters were verified with the single-shot emittance measurement and the spectrometer at the beginning of the PSB injection line. No difference in the emittance or energy spread was seen between the high current case and the normal operation beam.

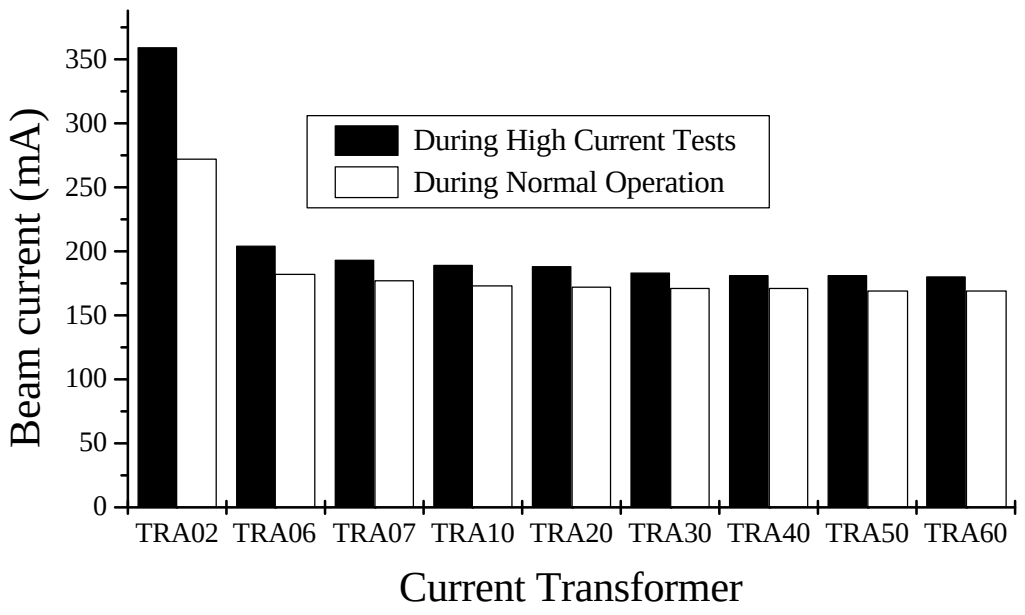

normal operation beam, and during tests of high current. TRA02:After Source, TRA06:after RFQ; TRA10: after accelerator; TRA20-60; transfer line to the PSB.

## 3 BUNCH SHAPE MONITORS

The principle of the Bunch Shape Monitor (constructed by INR, Troitsk) has been fully described in [3]. In short, a bunched ion beam impinges on a wire target held at high voltage, releasing secondary electrons that retain the initial bunch structure of the ion beam. The electrons are swept by a RF deflecting field, which allows the relative ion beam intensity at a given phase to be measured. By re-phasing the RF deflecting field with respect to the linac RF, the ion density distribution can be reconstituted.

On Linac 2, three BSMs are now installed (see Figure 2). The first two are standard devices placed in the inter-tank sections between tanks 1-2 and 2-3. At the output of the linac, the 3D-BSM allows selection of a transverse portion of the beam and the measurement of the bunch shape in a single-shot, by an array of charge collectors. The 3D-BSM was installed in 1996 [4], and the results of its first measurements on the proton beam are given in [5].

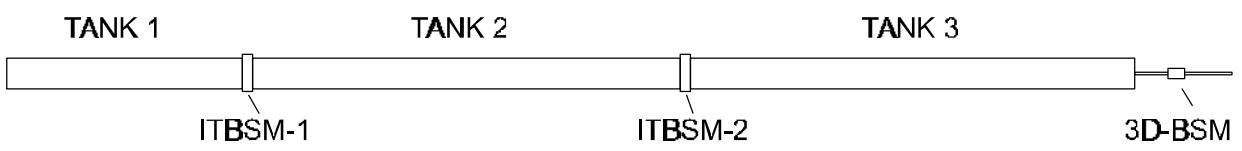

Figure 2. Scheme of the CERN 202 MHz Linac 2 Alvarez tanks, with the locations of the three BSMs.

The settings of the Alvarez tanks were studied by measuring the longitudinal movement of the bunch as a function of the RF phase and amplitude. By treating only the bunch centre, the effects of space charge in the calculation of the motion of the bunch are avoided. The bunch motion was calculated using the simulation tool DELTAT, based on the procedures given in [6].

The measurements were performed on the linac with the normal operational beam, in a time window from 25-50 µs after the start of the proton pulse from the source. Measurements have shown that before this time, the proton beam is not well stabilised (due to current variation and the time constant of the RF feedback loops.

## *2.1 Tank 1*

Measurements at the output of tank 1 have not been completed. The wire target of the IT-BSM is held at high voltage, but the current being drawn is too high for the HT supply (probably due to small discharges in vacuum). The rectification of this problem requires opening the vacuum of the Alvarez structure, which cannot be performed until the winter 2000-01 shutdown.

The comparison of the measured bunch position in phase and the simulation is complicated by the degrees of freedom (the unknown RF amplitude and the unknown offset in the bunch phase). The measured data are shown in Figure 3 along with curves from a simulation, for different RF levels of the tank. Note that the measured data can be arbitrarily offset vertically. The simulation results in the same gradient as the measurement, for RF tank levels of $0.95A_0$ , $1.10A_0$ and $1.15A_0$ (where $A_0$ is the nominal RF level). The power requirements to run the tank at 10% higher than the nominal values could not be fulfilled by the RF system, so the $0.95A_0$ line would be the most likely.

It is then estimated that the beam enters the tank with a phase of $-30^o$ during normal operation, whereas the initial synchronous phase of the tank is $-35^o$ (where $0^o$ is the crest of the RF wave).

With no second measurement of the beam of the bunch position as a function of the RF level, the results are not yet conclusive.

## *2.2 Tank 2*

With measurements of the bunch position in phase as a function of the tank 2 RF phase and amplitude, the comparison of the simulated and measured data is much easier. In Figure 4a the measured data are compared to the simulations using DELTAT. The gradient of the measured data is very similar to that of the results obtained for an input phase of $-10^o$, compared to the nominal synchronous phase of $-25^o$.

The bunch centre as a function of the tank 2 RF phase is given in Figure 4b, and is shifted along the x-axis such that the nominal phase corresponds to $-10^o$. The simulation fits the measured data well.

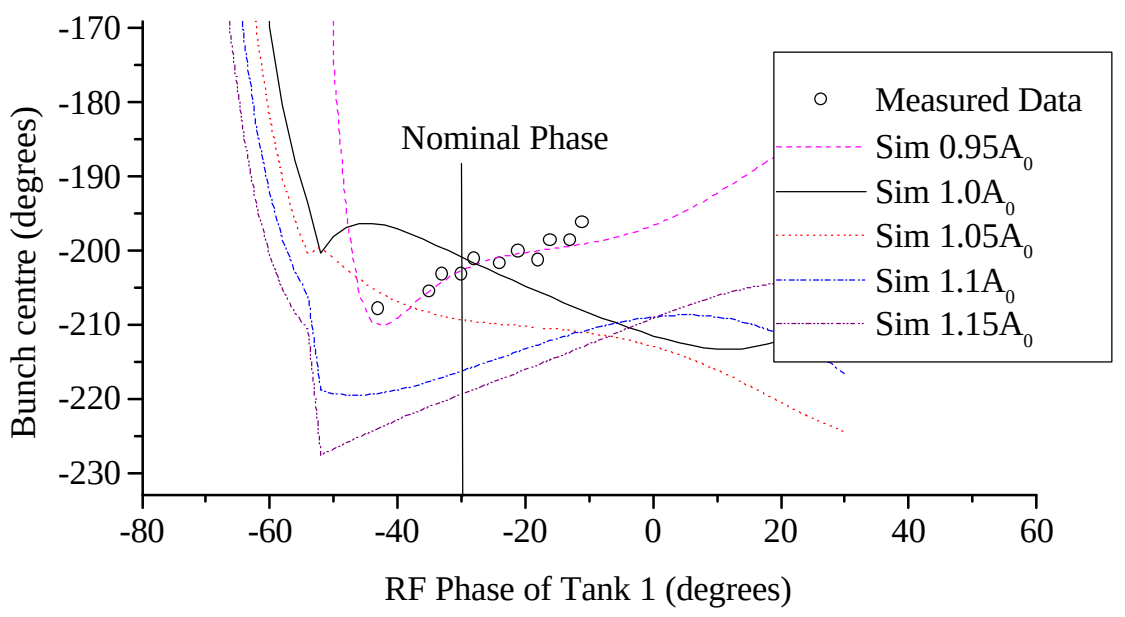

Figure 3. Measured data for the phase position of the bunch at the output of tank 1 as a function of the RF phase of tank 1. Curves show simulated bunch position as a function of the RF phase, for different RF amplitudes.

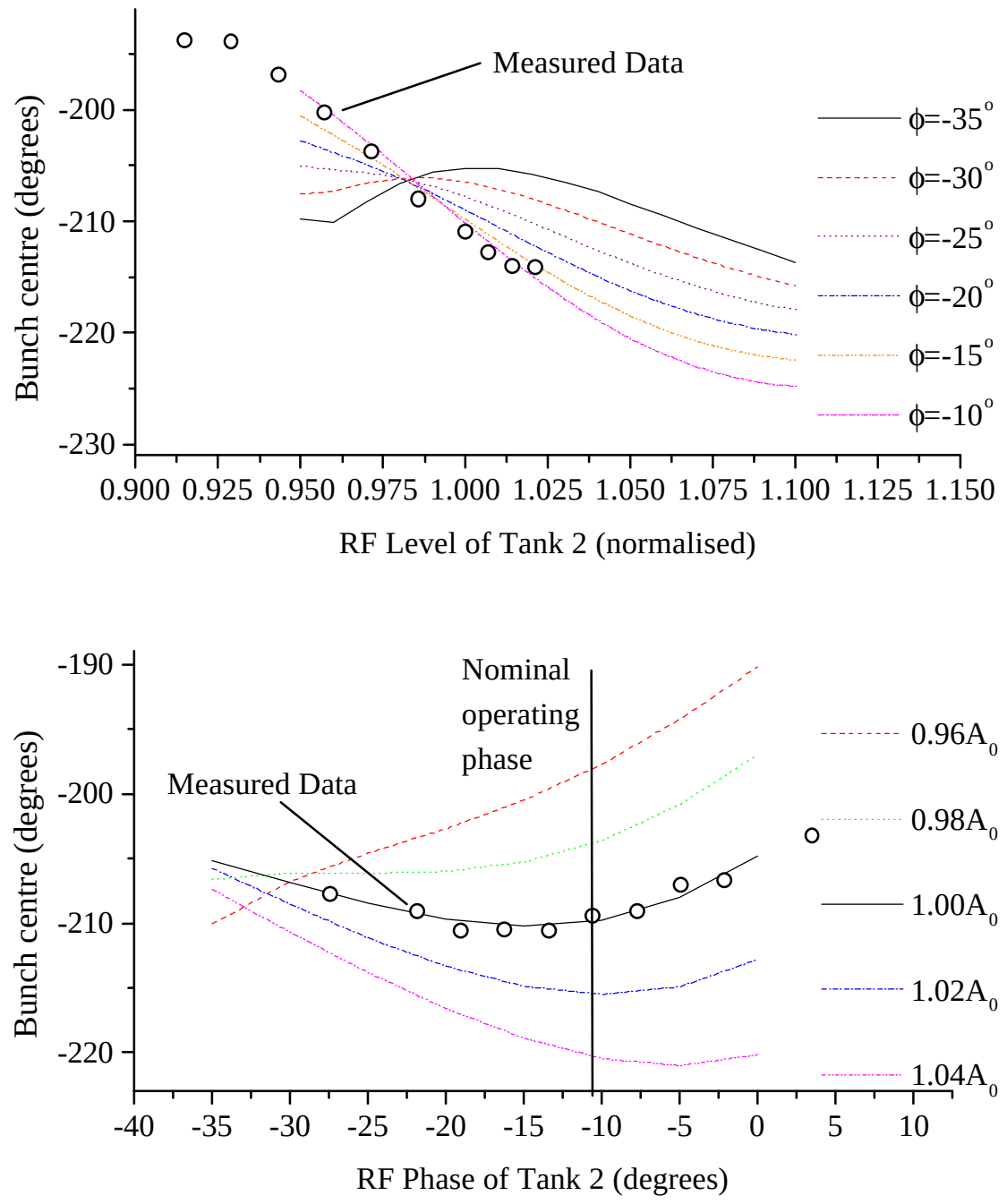

Figure 4. Measured data for the phase position of the bunch at the output of tank 2 as a function of a) the RF amplitude and b) the RF phase. Curves show simulated bunch phase position.

## *2.3 Tank 3*

The measurements with the 3D-BSM located after tank 3 have concentrated on the longitudinal dynamics, and the transverse distribution of the beam is not considered here. As the electron bunch is measured with an array of transverse charge collectors, the resolution during the measurements reported here is approximately $4.5^o$, which is 2 to 4 times lower than the resolution of the BSMs located between the tanks. Improving the resolution results in a narrower total range of phases that can be measured.

The measurements of the phase position of the bunch as a function of RF amplitude and phase are given in Figure 5. The simulations are in good agreement with the measured data based on an input phase of $-40^o$ compared to a nominal synchronous phase of $-25^o$, and with a RF field level 6% higher than the nominal value.

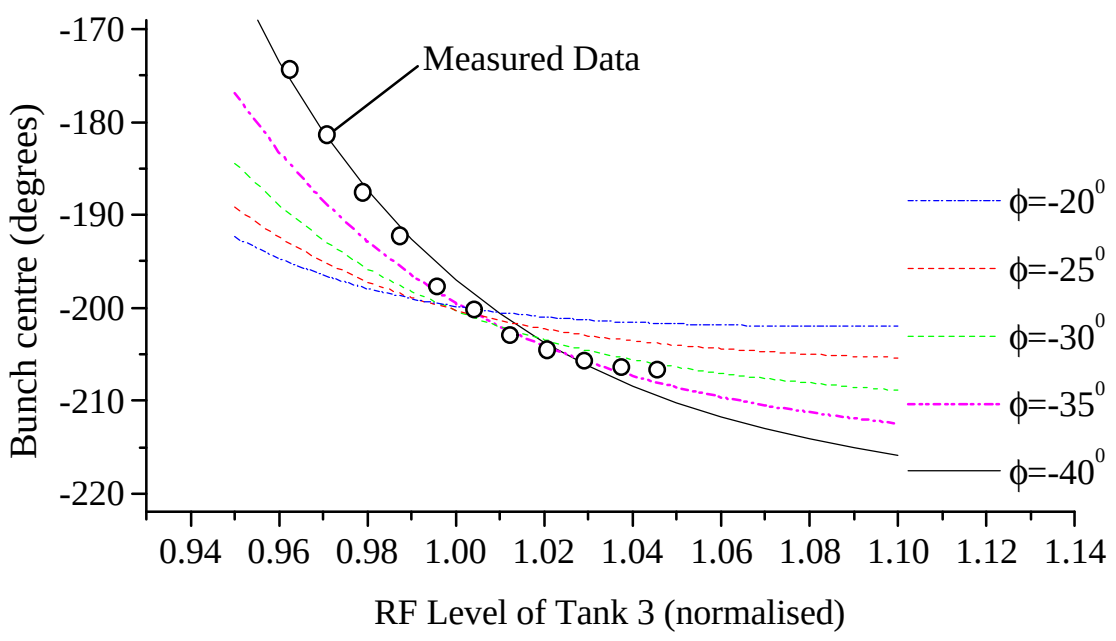

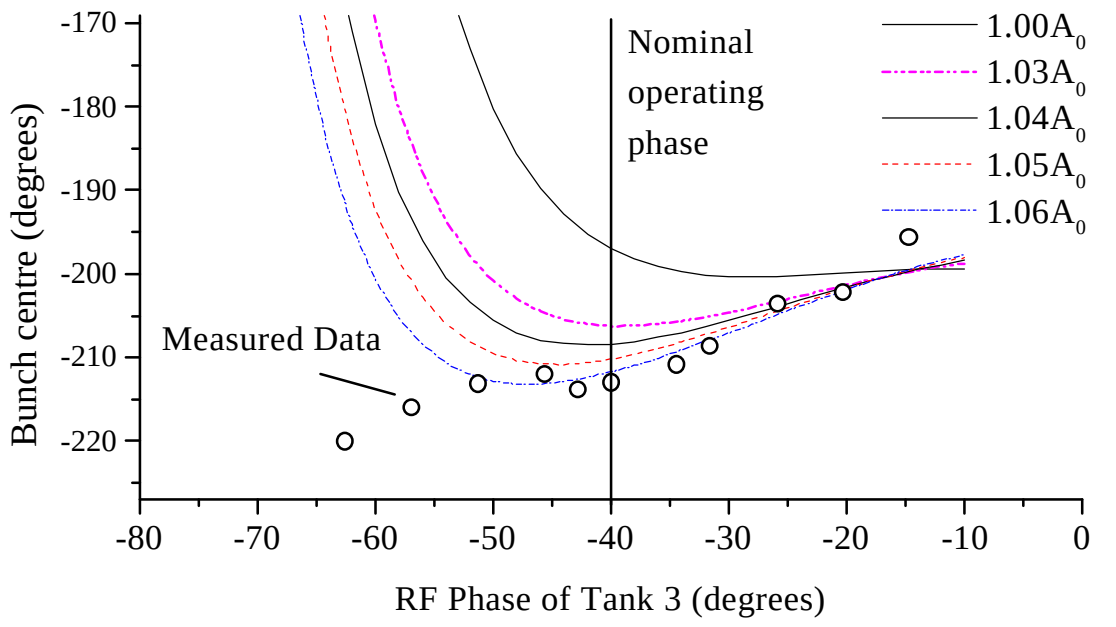

Figure 5. Measured data for the phase position of the bunch at the output of tank 3 as a function of a) the RF amplitude and b) the RF phase of Tank 3. Curves show simulated bunch position.

## 4 CONCLUSIONS

It has been demonstrated that the CERN lLinac 2 can provide the 180 mA beam required for the LHC injector chain. The larger resulting losses in this case mean that the beam is not at present used in routine operation at the PS complex.

The BSM data compiled for tank 1 are incomplete and require the repair of the BSM and the measurement of the bunch phase position as a function of RF amplitude before final conclusions can be drawn.

Tank 2 data are in excellent agreement with simulations of the bunch centre. Tank 3 data shows good agreement with simulations but with a RF level much higher than the nominal value.

This exercise provides an excellent starting point for further simulation with a macro-particle code (e.g. PARMILA) to provide more complete understanding of the dynamics and limitations of the structure with higher currents. This should allow the losses at higher currents to be reduced for routine operation as the LHC preinjector.